\documentclass{PoS}
\usepackage{amsmath,graphicx}

\newcommand{\cL}{{\cal L}}

\newcommand{\be}{\begin{equation}}
\newcommand{\ee}{\end{equation}}
\newcommand{\bea}{\begin{eqnarray}}
\newcommand{\eea}{\end{eqnarray}}

\newcommand{\hk}{\hspace{0.1cm}}

\newcommand{\rkx}{\right)}
\newcommand{\lk}{\left(}

\newcommand{\il}{\int\limits}

\newcommand{\vx}{\vec{x}}
\newcommand{\vy}{\vec{y}}

\newcommand{\vA}{\vec{A}}
\newcommand{\vE}{\vec{E}}
\newcommand{\vB}{\vec{B}}

\renewcommand{\vec}[1]{\mbox{\boldmath$#1$\unboldmath}}

\title{Hamiltonian approach to Yang-Mills theory in Coulomb
gauge}

\ShortTitle{Hamiltonian approach to Yang-Mills theory in Coulomb gauge}

\author{H. Reinhardt\thanks{Invited talk given by H. Reinhardt at the Workshop on QCD Green's
Functions, Confinement and Phenomenology, September 2009, Trento}\\
        Institut f\"ur Theoretische Physik\\
Auf der Morgenstelle 14\\
D-72076 T\"ubingen\\
Germany\\
        E-mail: \email{hugo.reinhardt@uni-tuebingen.de}}

\author{G.~Burgio $^a$, D.~R.~Campagnari $^a$, D.~Epple $^a$, C.~Feuchter $^a$, M.~Leder $^a$, M.~Pak $^{a}$, J.~M.~Pawlowski $^{b}$, M.~Quandt $^a$, W.~Schleifenbaum $^{a}$
        and A.~Weber $^{c}$\\
\llap{$^a$}T\"ubingen, Universit\"at,
     Institut f\"ur Theoretische Physik, Auf der Morgenstelle 14, D-72076
     T\"ubingen\\
\llap{$^b$}Heidelberg, Universit\"at,
       Institut f\"ur Theoretische Physik, Philosophenweg 16, D-69120 Heidelberg\\
\llap{$^c$}Morelia, Universidad Michoacana de San Nicol\'as de Hidalgo, Instituto de F{\'\i}sica y Matem\'aticas, Ciudad Universitaria, 58040 Morelia, Michoac\'an, Mexico\\
       }

\abstract{I review results recently obtained within the Hamiltonian approach to Yang-Mills
theory in Coulomb gauge. In particular, I will present results for the ghost and
gluon propagators and compare these with recent lattice data. Furthermore, I will
give an interpretation of the inverse of the ghost form factor as the dielectric
function of the Yang-Mills vacuum. Our ansatz for the vacuum wave functional
will be checked by means of functional renormalization group flow equations,
which are solved for the gluon energy and the ghost form factor. Finally, we
calculate the Wilson loop for the vacuum wave functional obtained from the
variational approach, using a Dyson equation.}

\FullConference{Workshop on QCD Green's
Functions, Confinement and Phenomenology, September 2009, Trento}

\begin{document}

\section{Introduction}

The usual language of quantum field theory is the functional integral
framework.  However, from ordinary quantum mechanics we know that for
many non-perturbative studies the solution of the Schr\"odinger
equation is much simpler than calculating the corresponding functional
integral. Although in quantum field theory the regularisation and
renormalisation procedures are much better understood in the
functional integral formalism, for non-perturbative investigations
like the infrared sector of Yang-Mills theory, the canonically
quantised operator formalism seems to have certain advantages over
more traditional field theoretical approaches if it comes to the
computation of physical observables. In recent years there have been
many activities in studying the infrared sector of QCD in Coulomb
gauge. The use of the Coulomb gauge is advantageous since this gauge
is a so-called ``physical gauge'': The gauge degrees of freedom can be
directly removed and Gauss' law can be explicitly resolved. For
example, the confining potential between static colour sources can be
extracted much more easily than e.g. in Landau gauge. In this talk I
report on results obtained in recent years by our group by a
variational solution of the Yang-Mills
Schr\"odinger equation in Coulomb gauge. The plan of my talk is as follows:\\
\noindent After these introductory remarks I will briefly review the basic ingredients of
the Hamiltonian approach to Yang-Mills theory in Coulomb gauge. In Sect. 3 I
will present some results obtained from a variational solution of the Yang-Mills
Schr\"odinger equation in Coulomb gauge and compare them with recent lattice
data. There I will give a phy\-sical inter\-pretation of the ghost form factor
in Coulomb gauge. Our variational ansatz for the Yang-Mills wave functional will
be checked in Sect. 4 by means of the functional renormalisation group flow
equations, which will be solved assuming ghost dominance. In Sect.\ 5 and 6 I will
present some applications of our approach. Thereby I will focus on the
calculation of the topological susceptibility and of the Wilson loop by means of a recently proposed Dyson equation. 

\section{Hamiltonian approach to Yang-Mills theory in Coulomb gauge}

Standard canonical quantisation of Yang-Mills theory in Weyl gauge $A_0 = 0$
leads to the Hamiltonian
\be
\label{g1}
H = \frac{1}{2} \int \lk \Pi^2 + B^2 \rkx \hk ,
\ee
where $\Pi = \delta / i \delta A$ is the momentum operator and $B^a_i$ the
non-Abelian magnetic field. Due to the Weyl gauge Gauss' law escapes from the
equation of motion and has to be implemented as a constraint on the wave
functional
\be
\label{g2}
\hat{D} \Pi | \psi \rangle = \rho | \psi \rangle \hk .
\ee 
Here $\hat{D} = \partial + g \hat{A} , \hat{A}^{ab} = f^{acb} A^c$ is the
covariant derivative in the adjoint representation of the gauge group and $\rho$
denotes the colour charge of the matter fields. Implementing the Coulomb gauge
$\vec{\partial} \vA = 0$ Gauss' law (\ref{g2}) can be explicitly resolved
resulting in the Coulomb gauge fixed Hamiltonian
\be
\label{g3}
H = \frac{1}{2} \int \lk J^{- 1} \Pi^\perp J \Pi^\perp + B^2 \rkx +
\frac{g^2}{2} \int J^{-1} \lk \rho + \rho_{dyn} \rkx J F (\rho + \rho_{dyn}) \hk ,
\ee
where $\Pi^\perp$ denotes the transversal momentum operator, $J = \mathrm{Det} (- \hat{D}
\partial)$ is the Faddeev-Popov determinant and $\rho^a_{dyn} = - \hat{A}^{ab}_k
\Pi^{b \perp}_k$ is the colour charge of the gauge field. Furthermore
\be
\label{g4}
F^{ab} (x, y) = \langle a, x | (- \hat{D} \partial)^{- 1} (- \partial^2) (-
\hat{D} \partial)^{- 1} |b, y \rangle
\ee
is the so-called Coulomb kernel. Its vacuum expectation value $\langle F
\rangle$ defines the non-Abelian Coulomb potential. In the gauge fixed theory the
matrix elements of an operator $O [A, \Pi]$ between states of the Yang-Mills
Hilbert space are defined by
\be
\label{g5}
\langle \psi_1 | O [A, \Pi] | \psi_2 \rangle = \int D A^\perp J [A^\perp] \psi^*_1
[A] O [A, \Pi] \psi_2 [A] \hk ,
\ee
where the restriction to the integration over the transversal gauge fields and
the Faddeev-Popov determinant in the integration measure arise from the
implementation of the Coulomb gauge.

In Ref.\ \cite{FeuRei04} a variational solution of the Yang-Mills Schr\"odinger equation
has been accomplished using the following ansatz for the vacuum wave
functional\footnote{In Refs.\ \cite{SzcSwa01}, \cite{Szc04} a pure Gaussian ansatz was used.
Furthermore, in Ref.\ \cite{SzcSwa01} the Faddeev-Popov determinant was completely ignored, while in Ref.\ \cite{Szc04} it was
included in the kinetic part of the Yang-Mills Hamiltonian
only. For a more general discussion of the ans\"atze for the vacuum wave functional, see Ref.\ \cite{Feuchter:2004gb}}
\be
\label{g6}
\psi [A] = J [A]^{- 1/2} \exp \left[ - \frac{1}{2} \int A \omega A \right] \hk ,
\ee
where $\omega (\vx, \vy)$ is a variational kernel, which is determined by
minimising the energy 
\be
\label{g7}
\langle H \rangle = \frac{\langle \psi | H | \psi \rangle}{\langle \psi | \psi
\rangle} \hk .
\ee
For the wave functional (\ref{g6}) the static gluon propagator is given by
\be
\label{g8}
\langle A A \rangle = (2 \omega)^{- 1} \hk 
\ee
implying that $\omega$ represents the gluon energy. In Ref.\ \cite{FeuRei04}
the energy $\langle H \rangle$ was calculated up to two loops.
Furthermore the gap equation $\delta \langle H \rangle / \delta \omega = 0$ was
converted into a set of Dyson-Schwinger equations.

\section{Results}

\begin{figure}
\centering
\includegraphics[height=4.5cm]{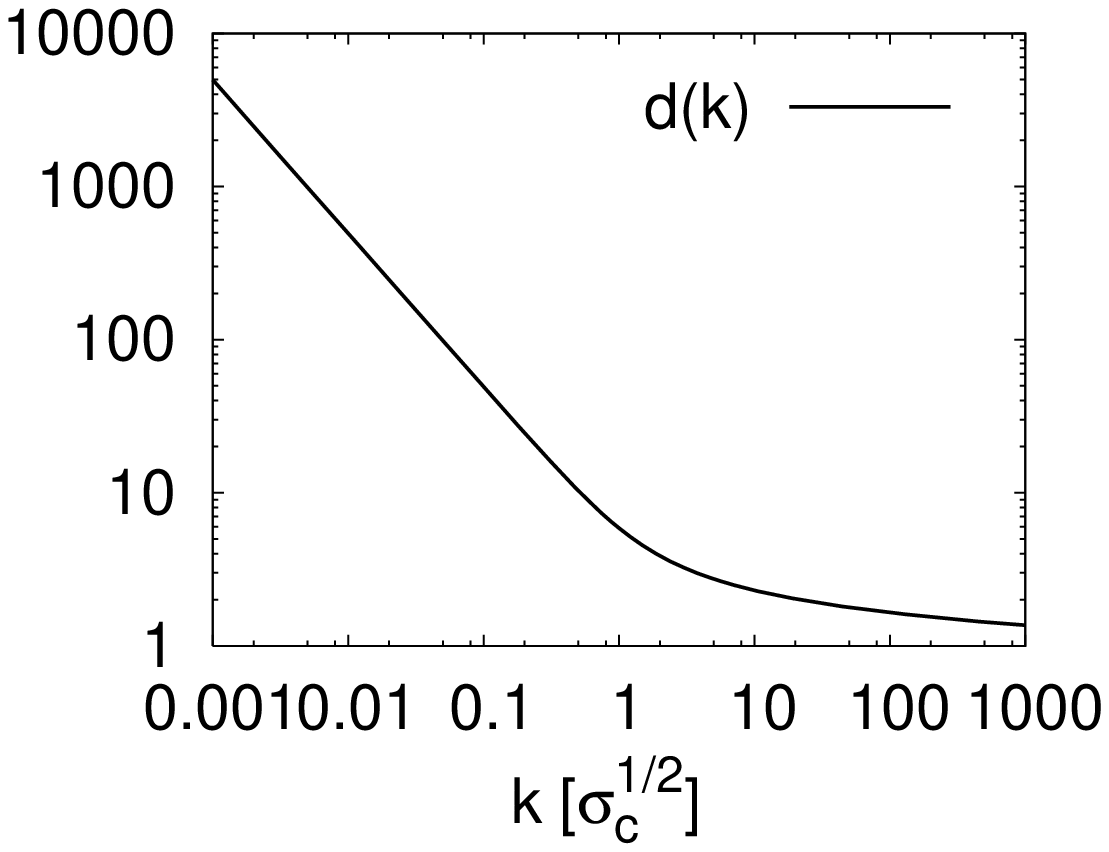}  \qquad \includegraphics[height=4.5cm]{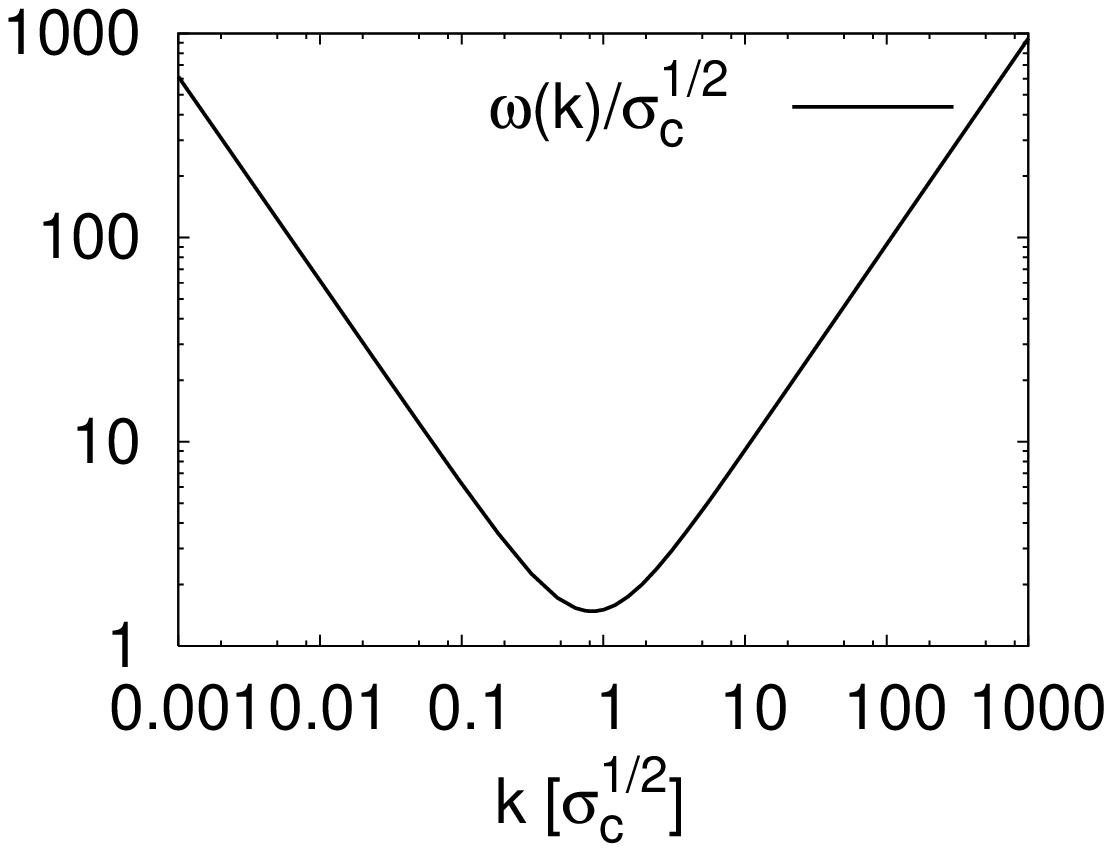} 
\caption{\sl Ghost form factor (left panel) and gluon energy $\omega(k)$ (right panel). }
\label{fig1}
\end{figure}

\begin{figure}
\centering
\includegraphics[height=4.5cm]{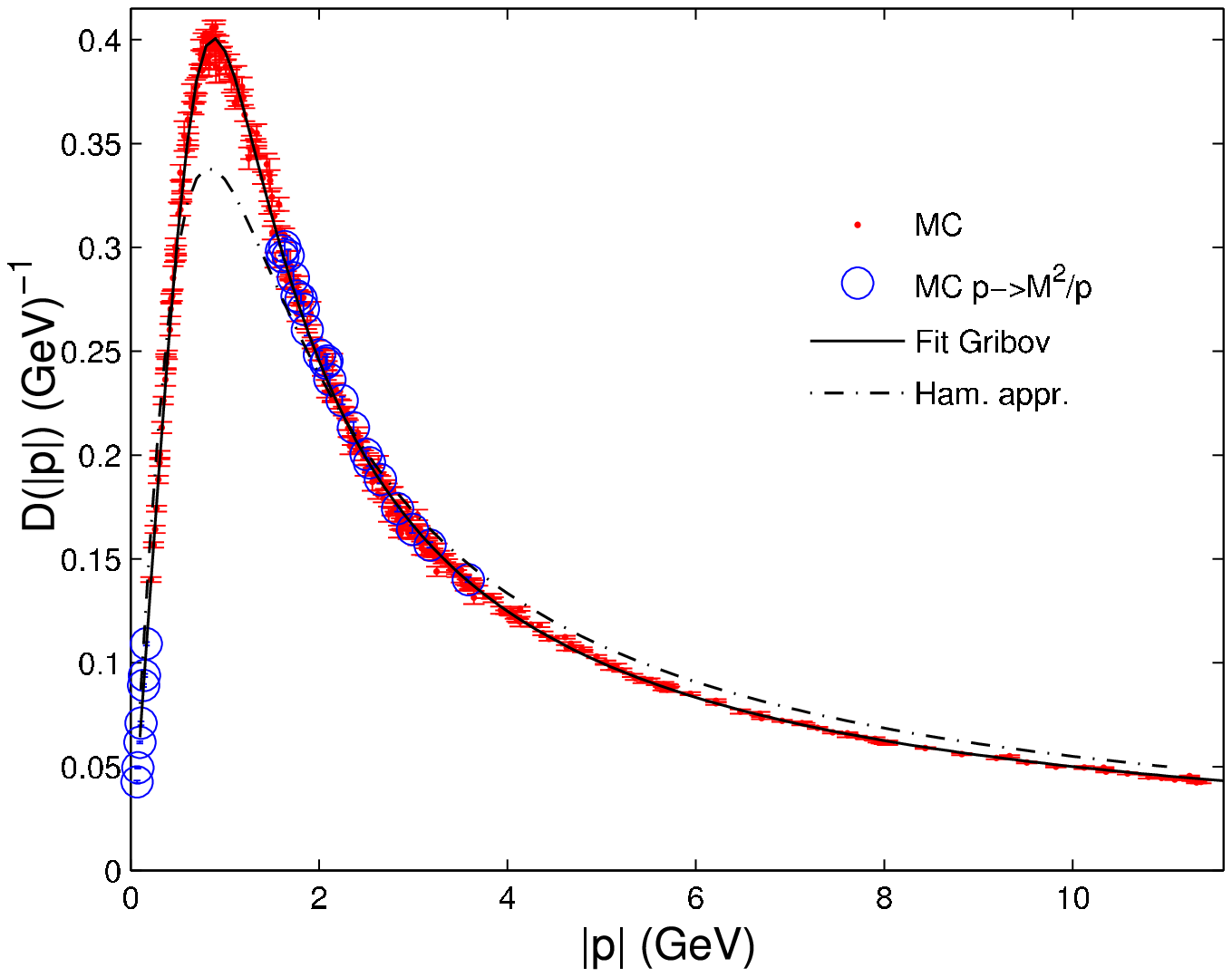} \qquad  \includegraphics[height=4.5cm]{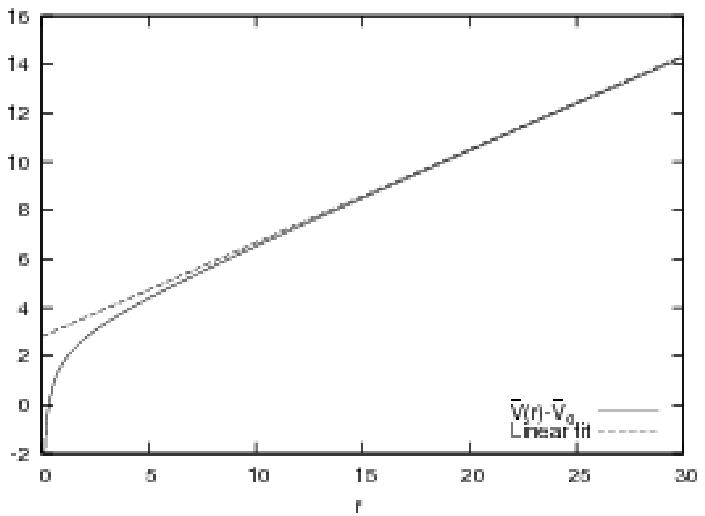} 
\caption{\sl Static gluon propagator $1/(2\omega(k))$
compared to the lattice results (left panel) and static quark potential (right panel).}
\label{fig3}
\end{figure}

\begin{figure}
\centering
\includegraphics[height=4.5cm]{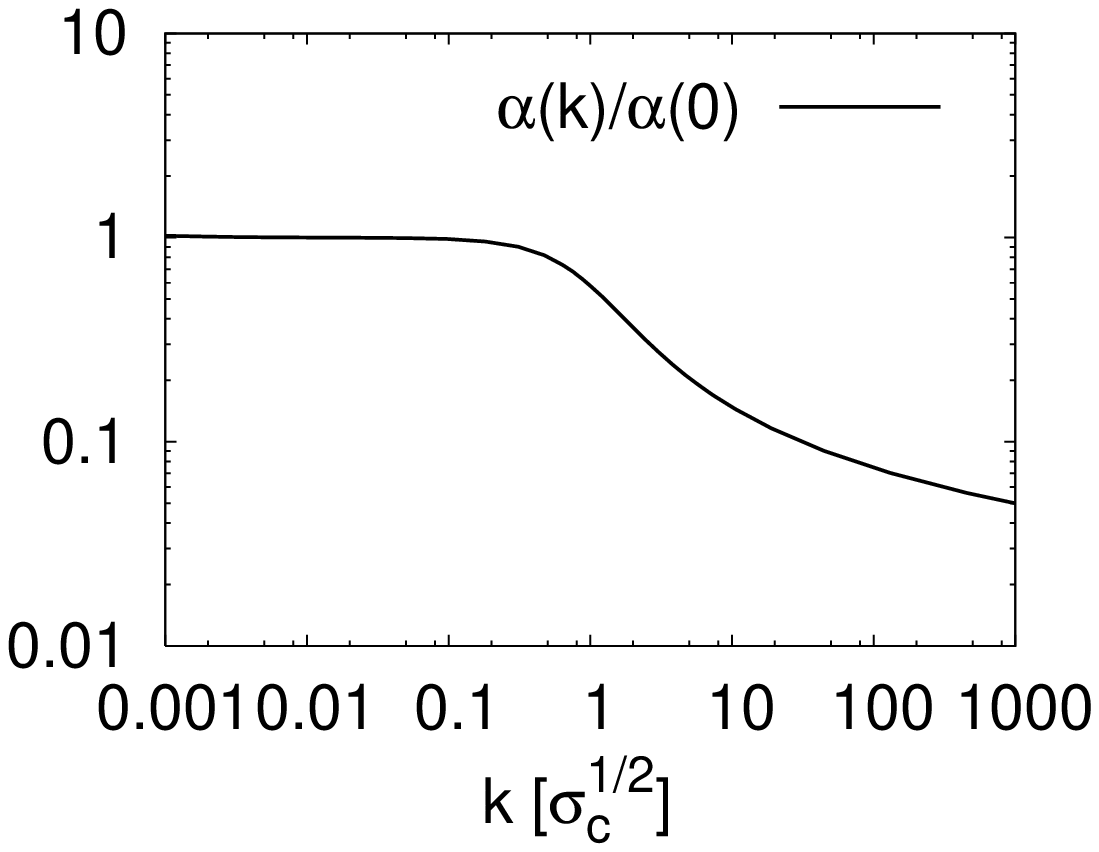} \qquad \includegraphics[height=4.5cm]{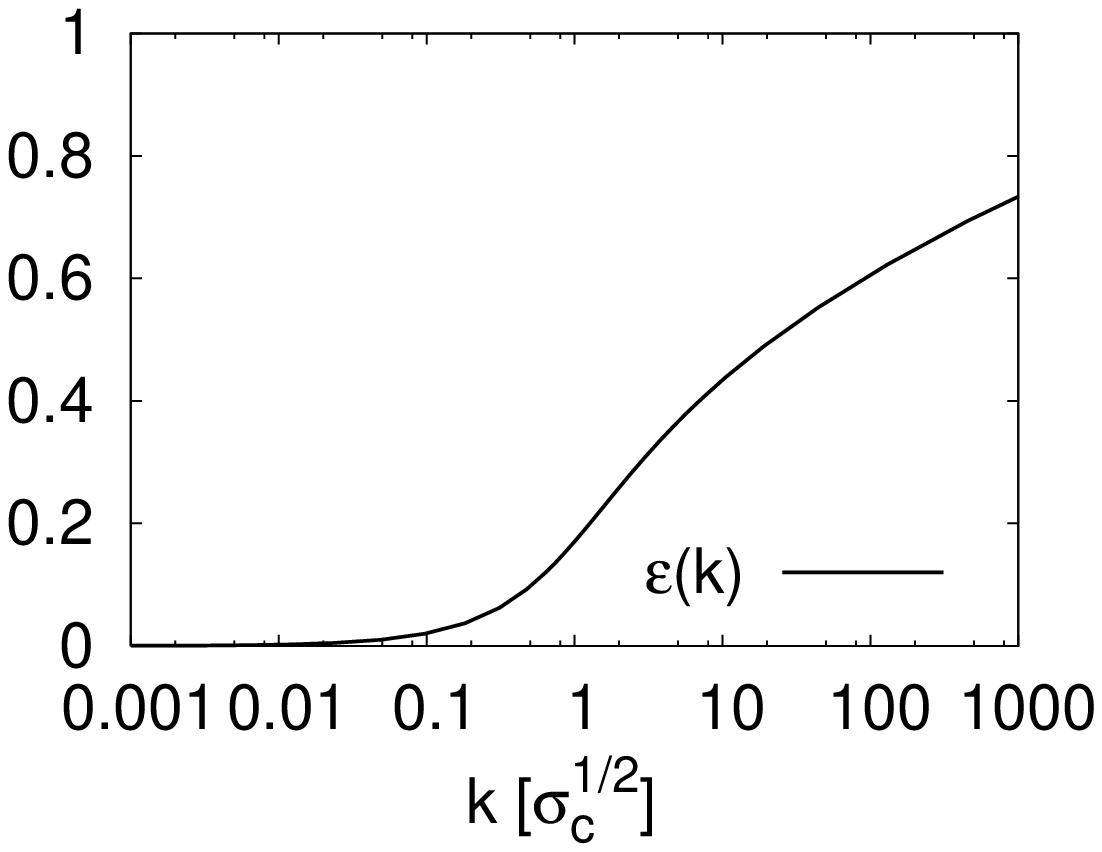}
\caption{\sl Running coupling constant (left panel) and dielectric function of the
Yang-Mills vacuum (right panel).}
\label{fig4}
\end{figure}

An infrared analysis, Ref.\ \cite{SchLedRei06}, of the Dyson-Schwinger equations shows that
the infrared exponents of the gluon and ghost propagators defined by 
\be
\label{g9}
\omega (k) \sim k^{- \alpha} , \quad d (k) = k^2 \langle (-\hat{D} \partial)^{-1} \rangle \sim k^{- \beta}
\ee
satisfy the sum rule
\be
\label{g10}
\alpha = 2 \beta - 1
\ee
and allow for two solutions
\begin{align}
\label{g11}
i)\quad  \beta \simeq 0.796 \, (0.85) \qquad
ii)\quad \beta = 1.0 \, (0.99) 
\end{align}
which were also produced by the numerical solutions obtained in Refs.\ \cite{FeuRei04} and
\cite{EppReiSch07}, respectively. The corresponding numerically obtained infrared exponents are given in
the brackets. In the numerical solution the horizon condition
\be
\label{g13}
d^{- 1} (k = 0) = 0
\ee
was explicitly built in.\footnote{In $D=3+1$ there are also so-called subcritical solutions \cite{Epp+07}, satisfying $d^{-1}(0) \neq 0$.} Figures \ref{fig1} and \ref{fig3} show the ghost form
factor and gluon energy as functions of the momentum for the solution $ii)$. At
large distances the gluon energy rises linearly with the momentum as expected
from asymptotic freedom, while it is infrared divergent at small momenta.
Solution $ii)$ gives rise to a strictly linearly rising static colour potential
shown in fig.\ \ref{fig3}. The running coupling constant extracted from the
ghost-gluon vertex obtained for the solution $ii)$ is shown in fig.\ \ref{fig4} (left).
It is the solution $ii)$ shown in figs.~\ref{fig1} and \ref{fig3}, which is also supported by the lattice data obtained
in Ref.\ \cite{BurQuaRei08}.
Remarkably, the lattice gluon energy can be perfectly fitted by Gribov's formula
\be
\label{g14}
\omega (k) = \sqrt{k^2 + \frac{M^2}{k^4}} 
\ee
with the energy scale $M \approx 0.8 \hk \mathrm{GeV}$. The lattice calculations carried out
in Ref.\ \cite{BurQuaRei08} differ from previous lattice calculations in two respects: the
residual gauge invariance left after implementing the Coulomb gauge has been
explicitly fixed and the scaling violations have been eliminated
giving rise to a strictly multiplicatively renormalisable static gluon propagator. For more
details we refer to Ref.\ \cite{BurQuaRei08}.

In Ref.\ \cite{Rei08} it was shown that the inverse of the Coulomb gauge ghost form
factor $d^{- 1} (k)$ can be interpreted as the dielectric function of the
Yang-Mills vacuum
\be
\label{g15}
\epsilon (k) = d^{- 1} (k) \hk ,
\ee
which is shown in fig.\ \ref{fig4} (right). By the horizon condition (\ref{g13}) the
dielectric constant vanishes in the infrared $\epsilon (k = 0) = 0$, implying
that the Yang-Mills vacuum is a perfect dia-electric medium, in which by Gauss'
law $\vec{\nabla} (\epsilon \vE) = \rho$ free colour charges cannot exist and
thus have to be confined in colourless states. This is precisely the picture
underlying the $M I T$ bag model, which assumes that $\epsilon = 0$ in the
vacuum while $\epsilon = 1$ inside the hadronic bags. The magnetic analogue of a
perfect dia-electric medium is a superconductor  for which the magnetic
permeability $\mu$ defined by $B = \mu H$ vanishes ($B$-magnetic field,
$H$-induction). In the usual notion of duality, meaning the interchange between
electric and magnetic fields and charges, a medium with $\epsilon = 0$ is a
dual superconductor. Therefore, the Gribov-Zwanziger confinement scenario
assuming $d^{- 1} (k = 0) = 0$ implies by the identity (\ref{g15}) the dual
Meissner effect in the infrared. 

\section{Functional renormalisation group flows}

Our ansatz for the Yang-Mills vacuum wave functional can be tested by
comparison with results from functional renormalisation group flows
(FRG). The basic idea of the hamiltonian FRG is to add an infrared
cut-off term quadratically in the quantum field $\phi$ \be
\label{g16}
\Delta S_k [\phi] = \frac{1}{2} \, \int d^3 p\, \phi (p) R_k (p) \phi (- p) \ee to
the Euclidean action, which cuts off momentum modes of the field
$\phi$ with momenta $p < k$, but leaves the theory unchanged for
momenta $p > k$. At large cut-off scales $k\to \Lambda_{UV}$ the theory
is well under control due to asymptotic freedom and perturbation
theory can be applied. In turn, for $k\to 0$ one recovers the full
theory. The flow of the theory with the cut-off scale $k$ is described
by the renormalisation group flow equation for the infrared
regularised effective action $\Gamma_k [\phi]$. For cut-off terms
(\ref{g16}) the flow equation reads \be
\label{g17}
\partial_t \Gamma_k [\phi] = \frac12 \, \mathrm{Tr} \left[ 
\left( \Gamma_k^{(2)}[\phi]+R_k \right)^{-1}\partial_t
  R_k \right]  ,
\qquad \partial_t \equiv k \, \frac{\partial}{\partial k} 
\ee 
where $\Gamma_k^{(2)}[\phi]+R_k $  is the inverse propagator of $\phi$ with 
\be
 \label{g18}
 \Gamma^{(2)}_k[\phi] = \frac{\delta^2 \Gamma_k[\phi]}{\delta \phi
   \delta \phi} \ee The flow equation (\ref{g17}) entails the
 evolution of the IR-regularised effective action from $k \to \infty$
 where $\Gamma_k [\phi]$ coincides with the bare action $S [\phi]$ to
 $k \to 0$ where $\Gamma_k [\phi]$ is the full effective action.
 
\begin{figure}
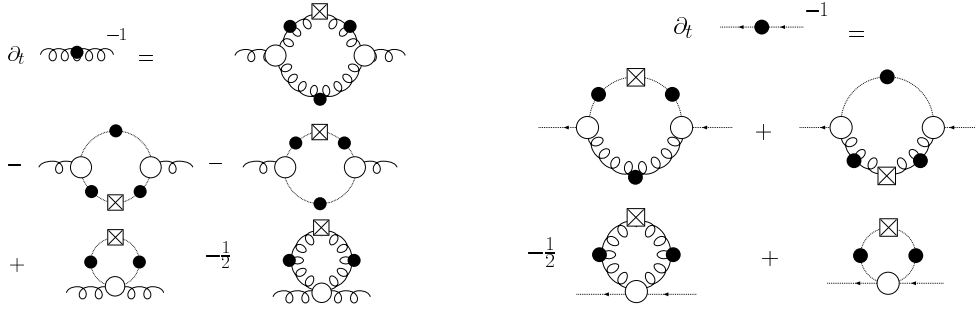

\begin{center}
\includegraphics[height=4cm]{fig8-1.epsi} \qquad\qquad \includegraphics[height=4cm]{fig8-2.epsi}
\caption{\sl FRG-flows for the propagators.}
\label{fig8}
\end{center}
\end{figure}

\begin{figure}
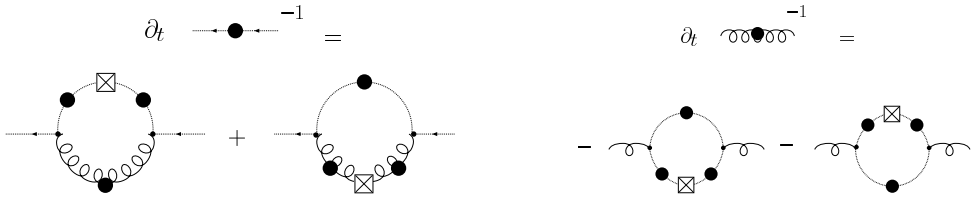

\begin{center}
\includegraphics[height=2.5cm]{fig9-1.epsi} \qquad\qquad \includegraphics[height=2.5cm]{fig9-2.epsi}
\caption{\sl FRG-flows for the propagators assuming ghost
dominance and dropping the tadpoles.}
\label{fig9}
\end{center}
\end{figure}

 Here we only are interested in the flow of the propagators, which is obtained
 from (\ref{g17}) by taking the second functional derivative with respect to the
 fields
 \be
 \label{g19}
 \partial_t \Gamma_k^{(2)} = 
\frac{\delta^2}{\delta\phi \, \delta\phi} \frac12 \, \mathrm{Tr} \left[ 
\left( \Gamma_k^{(2)}[\phi]+R_k \right)^{-1} \partial_t R_k \right] \biggr|_{\phi=0} \: .
 \ee
 This equation is diagrammatically illustrated in fig.\ \ref{fig8} 
for the propagators of Yang-Mills theory and is
 structurally similar
 to a Dyson-Schwinger equation except that the infrared regulator $\partial_t
 R_k$ enters the loops and all vertices and propagators are fully dressed. For the
 Hamiltonian flow of Yang-Mills theory in Coulomb gauge the fields involved are the
 transversal gauge field and the ghost field. Thus the right-hand side of the
 flow equation (\ref{g19}) receives contributions from ghost and gluon
 loops. We assume ghost dominance and keep only the contributions from the ghost
 loop to the right-hand side of the flow equation (\ref{g19}). The resulting
 flow equations for the gluon and ghost propagators are diagrammatically
 illustrated in fig.\ \ref{fig9}. With our choice of the vacuum wave functional
 the action is given by

\begin{figure}
\begin{center}
\includegraphics[height=4cm]{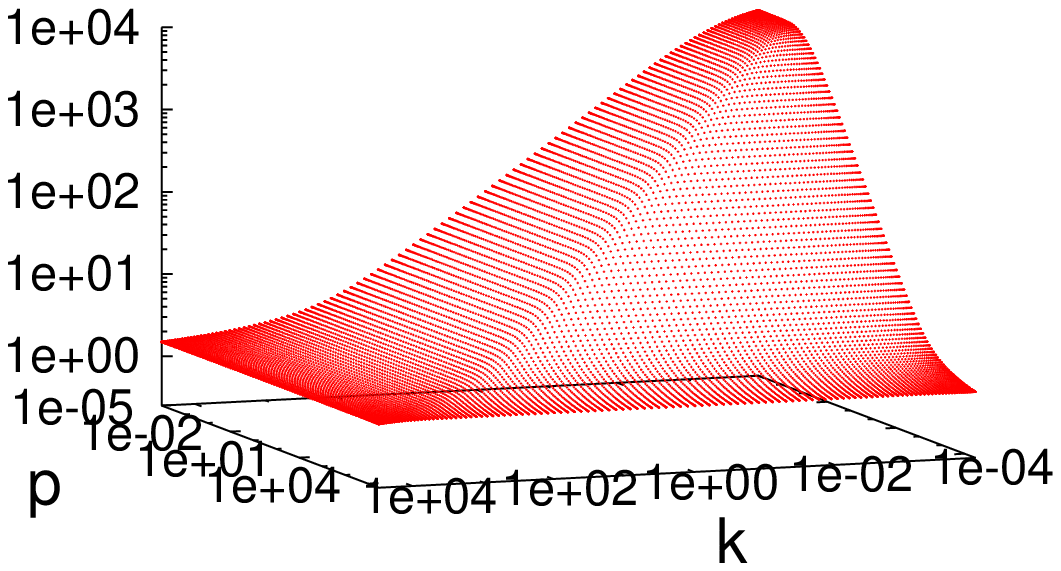} \qquad
\includegraphics[height=4.5cm]{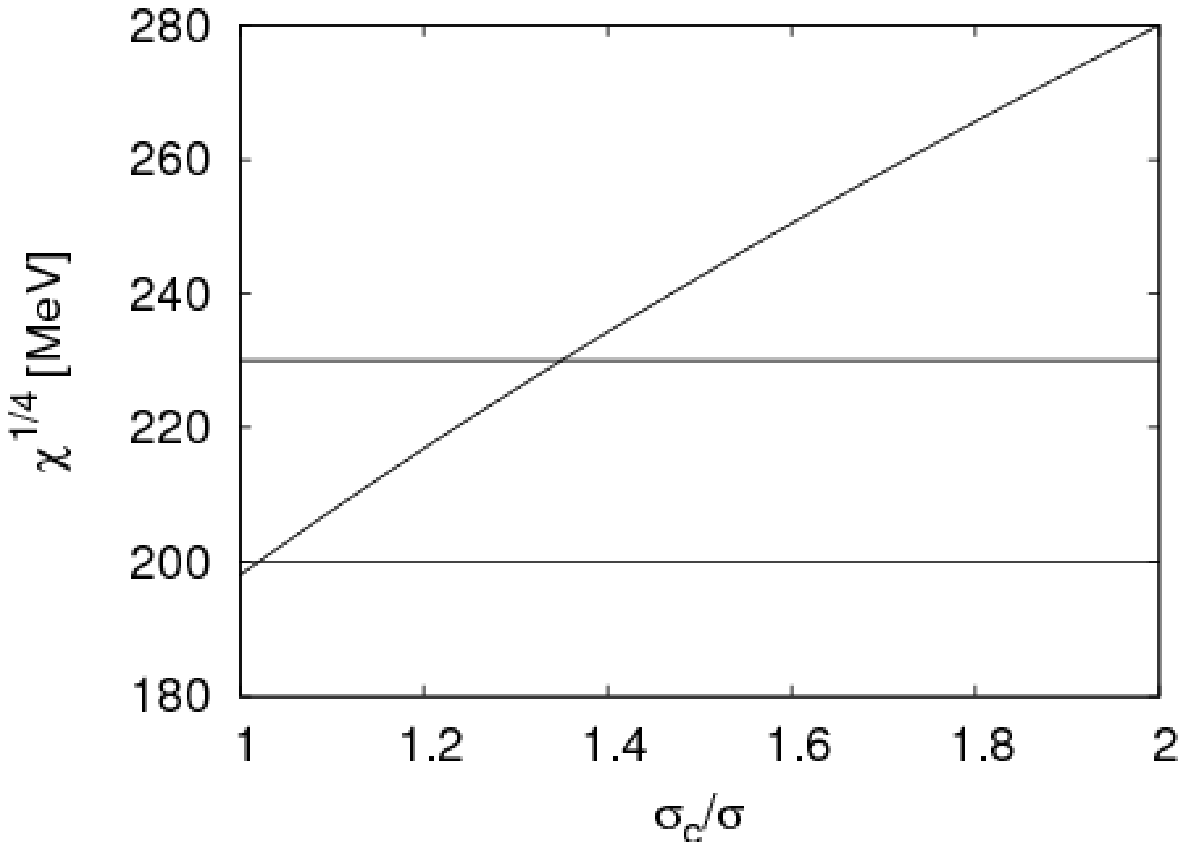}
\caption{\label{fig10} FRG-flow of the ghost form factor $d (p)$
  (left panel). Topological susceptibility obtained in the variational
  approach from solution $ii)$ (\protect\ref{g11}) as function of
  $\sigma_c/\sigma$ (right panel). The two horizontal lines limit the range of the lattice results.}\end{center}
\end{figure}
\begin{figure}
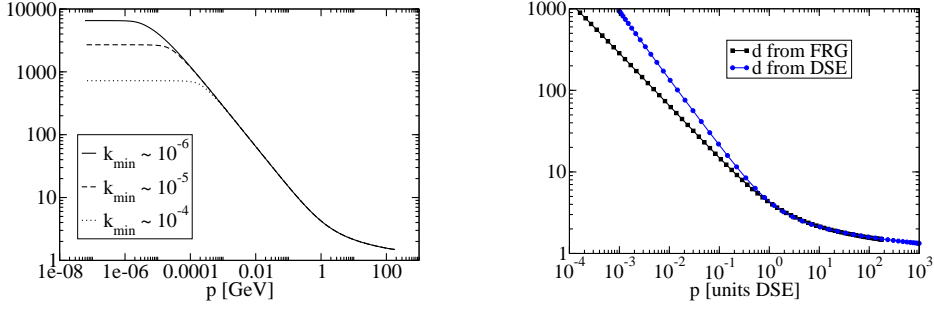

\begin{center}
\parbox{.4\linewidth}{\includegraphics[height=4cm]{fig11-2.eps}}
\qquad
\parbox{.4\linewidth}{\includegraphics[height=4cm]{fig11-1.eps}}
\caption{\label{fig11}\sl Ghost form factor $d_k(p)$ (left panel). For sake of comparison the
result of the variational approach is also shown (right panel).}\end{center}
\end{figure}
\begin{figure}
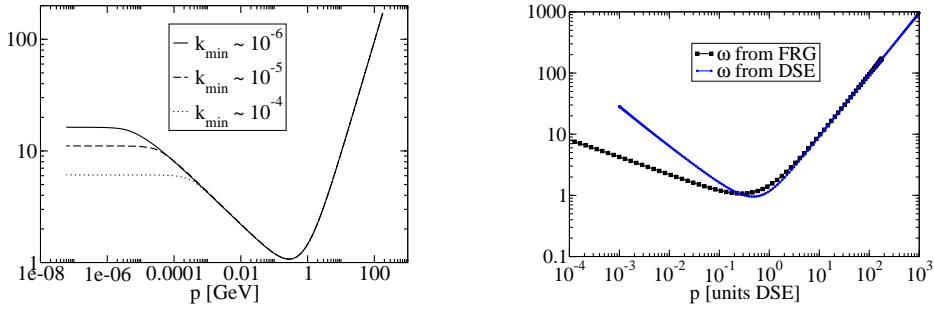

\begin{center}
\parbox{.4\linewidth}{\includegraphics[height=4cm]{fig12-2.eps}}
\qquad
\parbox{.4\linewidth}{\includegraphics[height=4cm]{fig12-1.eps}}
\caption{\label{fig12}\sl Gluon energy $\omega_k (p)$
obtained from the solution of the RG-flow equation (left panel). For sake of comparison the
result of the variational approach is also shown (right panel).}\end{center}
\end{figure}


 \be
 \label{g20}
 S [A, c, \bar{c}] = \int A \, \omega \, A + \int \bar{c} (-
 \hat{D} \partial) c \hk .  \ee We solve the FRG flows shown in fig.\
 \ref{fig9} for the gluon energy $\omega (p)$ and the ghost form
 factor $d (p)$ using the following regulators \bea
 \label{g21}
 \mbox{gluon:} \ R_{k} (p) \simeq p \:
 \exp\left(\frac{k^2}{p^2} -\frac{p^2}{k^2} \right) \; , \qquad
 \mbox{ghost:} \ R_k (p) \simeq p^2 \:\exp\left(\frac{k^2}{p^2} -\frac{p^2}{k^2} \right) \,, \hk \eea where we
 have suppressed the tensor structure, and perturbative initial
 conditions at large cut-off scale $k = \Lambda$ \bea
 \label{g22}
 \omega_\Lambda (p) = p + const \; , \qquad d_\Lambda (p) = const \hk
 \eea down to some minimal momentum cut-off scale $k_{min}$. Fig.\
 \ref{fig10} illustrates the renormalisation group flow of the ghost
 form factor. As the cut-off scale $k$ is reduced the ghost form
 factor as function of the momentum $p$ gets infrared enhanced and
 eventually becomes infrared divergent as the cut-off scale $k$
 becomes very small, i.e.\ the horizon condition emerges when the
 infrared cut-off is removed. This is nicely seen in fig.\ \ref{fig11}
 where the ghost form factor is shown at the minimum cut-off $k_{min}$
 as a function of the momentum $p$. It is also seen that the infrared
 exponent, i.e.\ the slope of the curve $d_{k_{min}} (p)$, does not
 change as the minimal cut-off scale is lowered. Fig.\ \ref{fig12}
 shows the corresponding result of the integration of the flow
 equation for the gluon energy $\omega_{k_{min}} (p)$. The infrared
 exponents of the gluon energy and the ghost form factor satisfy the
 sum rule (\ref{g10}) found from the Dyson-Schwinger equation of the
 variational approach. It can be shown that the FRG results coincide
 analytically with those of the variational approach either for
 optimised regulators \cite{Pawlowski:2005xe}, or if the ghost
 tadpoles are included. From analogous studies in the Landau gauge
 \cite{Pawlowski:2003hq} we expect that for general regulators the
 infrared exponents are smaller. Indeed this is the case for the
 approximation illustrated in fig.\ \ref{fig9} and with the regulators
 (\ref{g21}). This can be seen from figs.\ \ref{fig11} and
 \ref{fig12}. The gaps between the solutions in figs.\ \ref{fig11} and
 \ref{fig12} and the difference in the exponents gives an estimate for
 the systematic error of the present approximation. A detailed
 analysis will be provided elsewhere.

\section{Physical applications}

An important quantity for hadron physics is the topological susceptibility
$\chi$ defined by
 \be
 \label{g23}
\chi = \int d^4 x \: \langle 0 | q(x) \, q (0) | 0 \rangle \hk , \qquad  q (x) = \frac{g^2}{8\pi^2} \: \vE (x) \vB (x) \hk ,
  \ee
 where $q(x)$ is the topological charge density. It reflects the anomalous $U (1)$ symmetry
 breaking in QCD and can be extracted from the $\eta'$ mass through the
 Witten-Veneziano formula
  \be
 \label{g25}
 m^2_{\eta'} + m^2_{\eta} - 2 m^2_{K} = \frac{6}{F_\pi^2} \: \chi \hk , \qquad F_\pi = 93 \mbox{ MeV} .
 \ee
This quantity vanishes to all orders in perturbation theory and is therefore
perfectly suited to test non-perturbative methods. $\chi$ is a manifestation of
the $\theta-$vacuum and can be easily evaluated in the Hamiltonian approach.
Adding the topological term $\cL_{top} = \theta \int d^4x \; q(x)$ to the Yang-Mills Lagrangian shifts the momentum by $\vec{\Pi} \to \vec{\Pi} - \theta \, \frac{g^2}{8\pi^2} \, \vB$ and the Hamiltonian of the $\theta-$vacuum reads
 \be
 \label{g28}
 H_\theta = \frac{1}{2} \int \left[ \left(\Pi-\theta \, \frac{g^2}{8\pi^2} \, B \right)^2 + B^2 \right] ,
 \ee
from which one derives the following expression for the topological
susceptibility ($V$ is the spatial volume)

\begin{figure}
\originalTeX
\centerline{
\includegraphics[height=6cm]{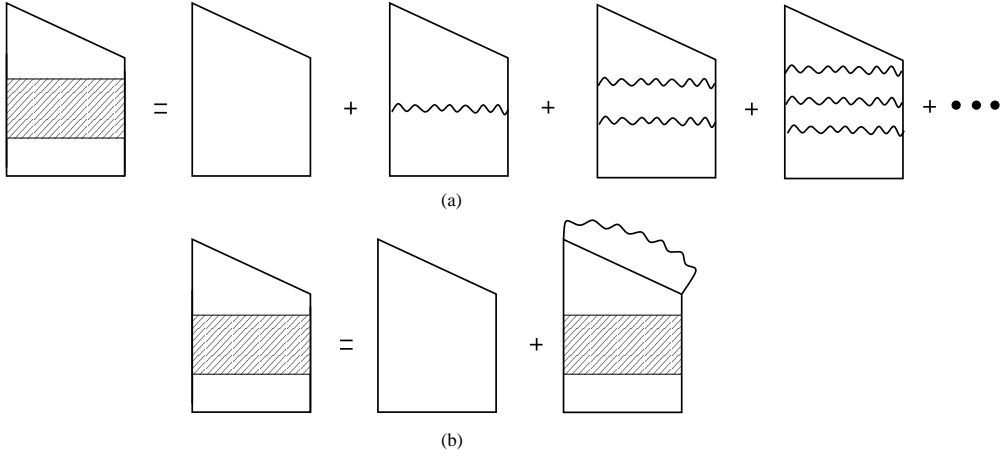}
}
\caption{\sl Graphical illustration to (a) the Dyson series for the Wilson loop and (b) the Dyson equation (\protect\ref{xyz-g1}).}
\label{fig15}
\end{figure}
 \be
 \label{g29}
 V \chi = \left. \frac{d^2 \langle H_\theta \rangle}{d \theta^2} \right|_{\theta
 = 0} \hk .
 \ee
 Using the Yang-Mills wave functional obtained from the variational principle
 for the solution $ii)$ (\ref{g11}) and restricting the intermediate states in
 eq.\ (\ref{g29}) to 3-gluon states (on top of the vacuum) one finds \cite{CamRei08} the results
 shown in fig.\ \ref{fig10} (right panel), where $\chi$ is plotted as function of the ratio
 $\sigma_c / \sigma$.  $\sigma_c$ is the Coulomb string tension, which is
 extracted from the static colour potential and used to fix the scale in the
 variational calculations while $\sigma$ is the Wilsonian string tension
 extracted from the Wilson loop. Lattice calculations indicate that this ratio is
 in the range of $\sigma_c / \sigma = 1.5.$ For such ratios the value obtained for
 $\chi$ is somewhat larger than the quenched lattice results.\\
 A crucial test for a wave functional would be the calculation of the Wilson
 loop, the order parameter of the Yang-Mills theory. The spatial Wilson loop can
 in principle be directly calculated in the Hamiltonian approach, once the vacuum
 wave functional is known. However, the practical calculation is rendered
 difficult by the path ordering. A quantity more easily accessible is the 't Hooft
 loop, the disorder parameter of Yang-Mills theory. In Ref.\ \cite{ReiEpp07} the 't Hooft
 loop was calculated using the wave functional corresponding to the solution
 $ii)$ (\ref{g11}) and a perimeter law was found, which is characteristic of
 the confinement phase. For further details we refer the reader to Ref.\ \cite{ReiEpp07}.

\section{Wilson loop from a Dyson equation}

Although the temporal Wilson loop is dual to the spatial t'Hooft loop and a perimeter law in the latter implies an area law in the former, one would appreciate an explicit calculation of the Wilson loop and observe the emergence of the area law. Recently, a Dyson equation has been proposed in the context of supersymmetric Yang-Mills theory, which, at least in an approximate fashion, takes care of the path ordering \cite{Erickson:1999qv}. It has been applied to the temporal Wilson loop in non-supersymmetric Yang-Mills theory \cite{Zayakin:2009jz}.\\
\noindent This Dyson equation applies to trapezoidal loops $W \left( S, T; L \right)$ and basically sums up the ladder diagrams shown in fig.~\ref{fig15}(a), i.e. the gluon exchange between one pair of opposite paths. This Dyson equation is diagrammatically shown in fig.~\ref{fig15}(b) and is analytically given by 
\be
\label{xyz-g1}
W (S, T; L) = 1 + g^2 C_2 \il^S_0 d s \il^T_0 d t D \left( \left( x (s) - s (t) 
\right)^2 \right) W (s, t; L)  \hk .
\ee
\begin{figure}
\originalTeX
\centerline{
\includegraphics[height=3cm]{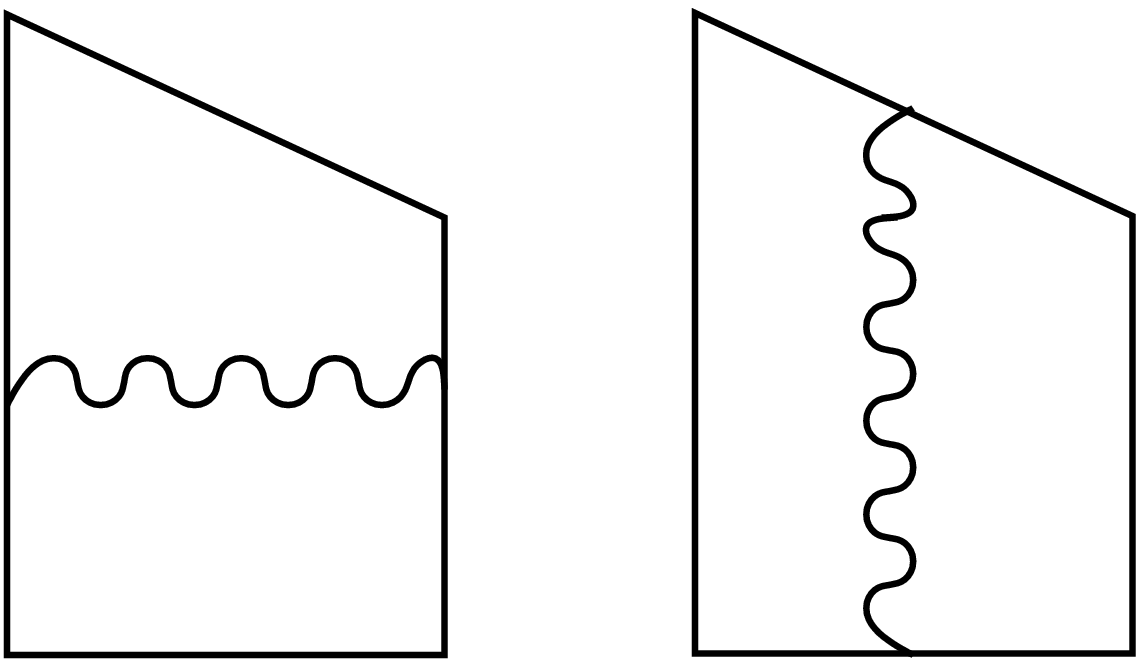}
\put(-35,-15){$(b)$}
\put(-125,-15){$(a)$}
}
\caption{\sl (a) Processes involved and (b) processes ignored in the Dyson equation (\protect\ref{xyz-g1}).}
\label{fig17}
\end{figure}

This equation suffers from the following limitations \cite{Pak:2009em}:
\begin{enumerate}
\item 
Since we include only one pair of paths (see fig.~\ref{fig17}(a)) the Dyson equation~(\ref{xyz-g1}) is restricted to strongly asymmetric loops. However, it is irrelevant whether the loop is a temporal or spatial one.
\item
When one of the two parallel temporal lines is set to zero, the trapezoidal loop degenerates to a triangle-shaped loop. In this case the Dyson equation~(\ref{xyz-g1}) yields for the Wilson loop 
\be
\label{xyz-g2}
W \left( S, T = 0; L \right) = 1
\ee
which, in general, is certainly incorrect. This wrong boundary condition is not surprising, since setting $T=0$ or $S=0$ contradicts the assumption $S,T \gg L$ in the Dyson equation~(\ref{xyz-g1}).
\item
The equation is bounded to produce strict Casimir scaling, which is known to occur only in the intermediate distance regime. 
\item
The Wilson loop is a gauge-invariant object. However, the right-hand side of the Dyson equation~(\ref{xyz-g1}) depends via the gluon propagator on the gauge. In fact, the gluon propagator can be defined only in a gauge-fixed theory and vanishes if the gauge is unfixed.

\item
Finally, the Wilson loop is renormalisation group invariant, while the right-hand side of the Dyson equation~(\ref{xyz-g1}) is not, except for the temporal Wilson loops in Coulomb gauge (which we will consider at first). 
\end{enumerate}
In Coulomb gauge the temporal gluon propagator has the form 
\begin{figure}
\begin{center}
\parbox{.4\linewidth}{\includegraphics[width=5.5cm,angle=-90]{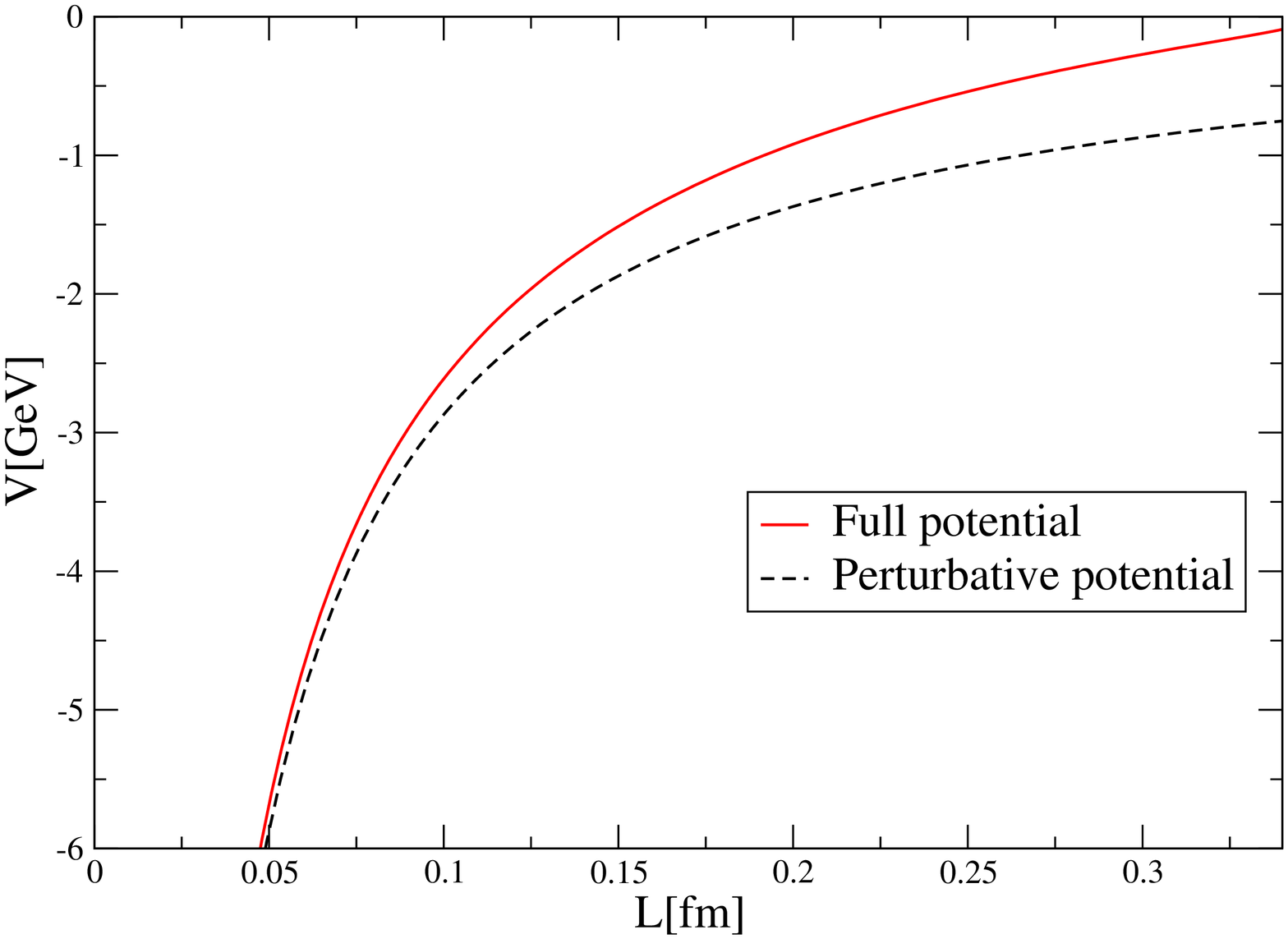}}
\qquad
\parbox{.4\linewidth}{\includegraphics[width=5.5cm,angle=-90]{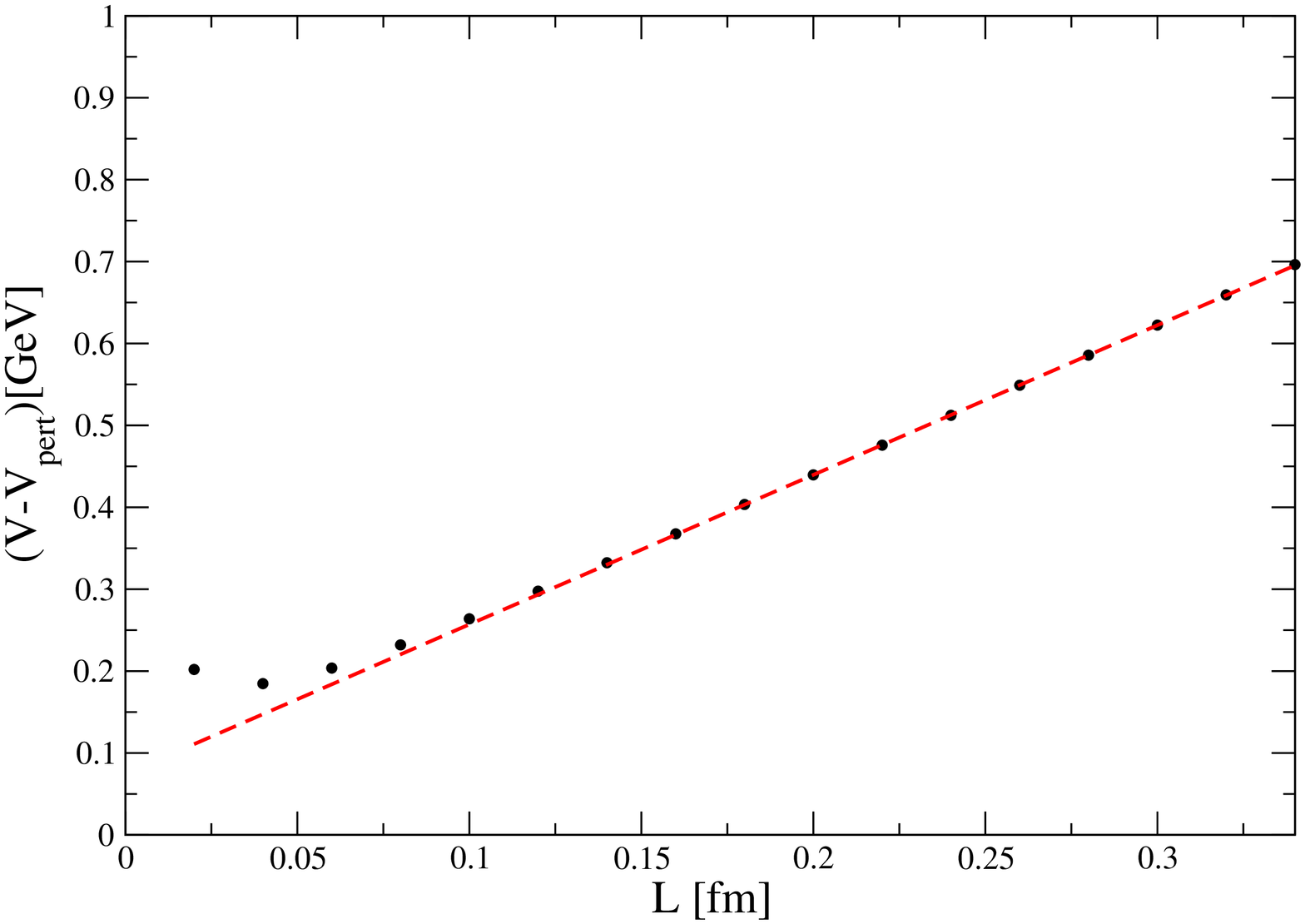}}
\caption{\label{fig16}\sl Left panel: The Wilsonian potential $V(L)$ obtained from the static gluon propagator (\protect\ref{g14}) and the perturbative potential $V_{\text{pert}}(L)$. Right panel: The full potential minus its perturbative part.}\end{center}
\end{figure}
\be
\label{xyz-g3}
g^2 D^{ab}_{00} (x, y) = - \delta^{ab} V_C (|\vx - \vy|) \delta (x^0 - y^0) +
P^{a b}
(x, y) \hk 
\ee
where $V_C (|\vx - \vy|)$ is the so-called non-Abelian Coulomb potential. At small distances it has the ordinary Coulombic $\sim 1/r$ behaviour, while it rises linearly at large distances $\sim \sigma_c r$, with $\sigma_c$ being the Coulomb string tension, which is larger than the Wilsonian string tension $\sigma$. The non-instantaneous part $P(x,y)$ is assumed to lower $\sigma_c$ towards $\sigma$.
If we ignore the screening part $P(x,y)$, the gluon propagator is instantaneous and the Dyson equation~(\ref{xyz-g1}) applies only to rectangular loops $\overline{W} \left( T; L \right) \equiv W \left( S = T, T; L \right)$ 
\be
\label{xyz-g5}
\overline{W} (T;L) = 1 - C_2 V_C (L) \il^T_0 d t \overline{W} (t;L) \hk .
\ee
This equation contains the correct boundary condition $\overline{W} \left( T = 0; L \right) = 1$ and can be solved analytically, yielding 
\be
\label{xyz-g6}
\overline{W} (T; L) = \exp \left( - C_2 V_C (L) T \right) \hk .
\ee
Within the approximation used for the gluon propagator, we have correctly obtained an area law. It is clear why in this case the equation produces the correct result: The processes neglected in the Dyson equation~(\ref{xyz-g1}) (see fig.~(\ref{fig17}(b)) do not exist for an instantaneous gluon propagator.\\
\noindent For arbitrary (non-instantaneous) gluon propagators, the Dyson equation~(\ref{xyz-g1}) can be converted into a one-dimensional Schr\"odinger equation 
\be
\label{xyz-g7}
\left[ - \frac{d^2}{d r^2} + U (r) \right] \varphi_n (r) = -
\frac{\Omega^2_n}{4} \varphi_n (r) \hk 
\ee
with the variable $r = \frac{S - T}{L}$ and the Schr\"odinger potential given by 
\be
\label{xyz-g8}
U (r) = - g^2 C_2 \, L^2 \, D \left( L^2 (1 + r^2) \right) \hk .
\ee
The Wilsonian potential can be obtained from the "ground state energy"
\be
\label{xyz-g9}
V \left( L \right) = - \lim_{T \to \infty} \, \frac{1}{T} \ln W \left( T, T; L \right) = - \frac{\Omega_0 \left( L \right)}{L} + const \; . 
\ee
Applying the Dyson equation to the spatial Wilson loop in Coulomb gauge and
using the static transversal gluon propagator (\ref{g14}) augmented by its
anomalous dimension (derived in Ref.\ \cite{Mexiko}) to make the gluon propagator well-defined in coordinate space, one finds the Wilsonian potential shown in fig.~\ref{fig16}(a). When one subtracts from this potential the perturbative one, a linearly rising potential is left, implying an area law for the spatial Wilson loop, see fig.~\ref{fig16}(b). 

\section*{Acknowledgements}

This work was supported in part by DFG under contract DFG-Re856/6-3, by DAAD, Conacyt grant 46513-F, CIC-UMSNH, by the
Europ\"aisches Graduiertenkolleg Basel-Graz-T\"ubingen and by the Cusanuswerk. JMP is supported by Helmholtz Alliance
HA216/EMMI.


\begin{thebibliography}{99}

\bibitem{FeuRei04} C.~Feuchter and H.~Reinhardt, 
Phys.\ Rev.\ \textbf{D70}, 105021 (2004); \texttt{[hep-th/0408236]}.\\
C.~Feuchter and H.~Reinhardt,
  \texttt{[hep-th/0402106].}

\bibitem{SzcSwa01} A.~P. Szczepaniak and E.~S. Swanson,
Phys.\ Rev.\ \textbf{D65}, 025012 (2001); \texttt{[hep-ph/0107078]}.

\bibitem{Szc04} 
A.~P.~Szczepaniak, 
Phys.\ Rev.\ \textbf{D69}, 074031 (2004); \texttt{[hep-ph/0306030]}.

\bibitem{Feuchter:2004gb}
 H.~Reinhardt and C.~Feuchter,
  Phys.\ Rev.\  D {\textbf 71}, 105002 (2005);
  \texttt{[hep-th/0408237]}.

\bibitem{SchLedRei06} W.~Schleifenbaum, M.~Leder, and H.~Reinhardt, 
Phys.\ Rev.\ \textbf{D73}, 125019 (2006); \texttt{[hep-th/0605115]}.

\bibitem{EppReiSch07} D.~Epple, H.~Reinhardt, and W.~Schleifenbaum,
Phys.\ Rev.\ \textbf{D75}, 045011 (2007); \texttt{[hep-th/0612241]}.

\bibitem{Epp+07} D.~Epple, H.~Reinhardt, W.~Schleifenbaum, and A.~P. Szczepaniak, 
Phys.\ Rev.\ \textbf{D77}, 085007 (2008); \texttt{[hep-th/0712.3694]}.

\bibitem{BurQuaRei08} G.~Burgio, M.~Quandt and H.~Reinhardt, 
Phys.\ Rev.\ Lett.\ \textbf{102} (2009) 032002; \texttt{[hep-lat/0807.3291]}.

\bibitem{Rei08} H.~Reinhardt,
Phys.\ Rev.\ Lett.\ \textbf{101} (2008) 061602; \texttt{[hep-th/0803.0504]}.

\bibitem{Pawlowski:2005xe}
  J.~M.~Pawlowski,
  Annals Phys.\  {\bf 322} (2007) 2831
  \texttt{[hep-th/0512261]}; \\
  D.~F.~Litim,
  Phys.\ Lett.\  B {\bf 486} (2000) 92
\texttt{[hep-th/0005245]}.

\bibitem{Pawlowski:2003hq} 
  J.~M.~Pawlowski, D.~F.~Litim, S.~Nedelko and L.~von Smekal,
    Phys.\ Rev.\ Lett.\  \textbf{93} (2004) 152002;
  \texttt{[hep-th/0312324]}.

\bibitem{CamRei08} D.~R.~Campagnari and H.~Reinhardt, 
Phys.\ Rev.\ \textbf{D78} (2008) 085001; \texttt{[hep-th/0807.1195]}.

\bibitem{ReiEpp07} H.~Reinhardt and D.~Epple, 
Phys.\ Rev.\ \textbf{D76}, 065015 (2007); \texttt{[hep-th/0706.0175]}.

\bibitem{Erickson:1999qv} J.~K.~Erickson, G.~W.~Semenoff, R.~J.~Szabo and K.~Zarembo,
  Phys.\ Rev.\ {\bf D61}, 105006 (2000);
  \texttt{[hep-th/9911088]}.\\
  J.~K.~Erickson, G.~W.~Semenoff and K.~Zarembo,
  Nucl.\ Phys.\ {\bf B582}, 155 (2000);
 \texttt{[hep-th/0003055]}.

\bibitem{Zayakin:2009jz}
    A.~V.~Zayakin and J.~Rafelski,
  Phys.\ Rev.\ {\bf D80}, 034024 (2009);
  \texttt{[hep-ph/0905.2317]}.

\bibitem{Pak:2009em}
  M.~Pak and H.~Reinhardt,
  \texttt{[hep-th/0910.2916]}.

\bibitem{Mexiko} 
D.~Campagnari, A.~Weber, H.~Reinhardt, F.~Astorga and W.~Schleifenbaum, 
\texttt{[hep-th/0910.4548]}. 

\end{thebibliography}
\end{document}